# Entropy of Pseudo Random Number Generators

Stephan Mertens[*] and Heiko Bauke[†]
*Institut für Theoretische Physik, Otto-von-Guericke Universität, PF 4120, 39016 Magdeburg, Germany*
(Dated: June 30, 2004)

Since the work of Ferrenberg *et al.* [PRL 69, (1992)] some pseudo random number generators are known to yield wrong results in cluster Monte Carlo simulations. In this contribution the fundamental mechanism behind this failure is discussed. Almost all random number generators calculate a new pseudo random number $x_i$ from preceding values, $x_i = f(x_{i-1}, x_{i-2}, \ldots, x_{i-q})$. Failure of these generators in cluster Monte Carlo simulations and related experiments can be attributed to the low entropy of the production rule $f()$ conditioned on the statistics of the input values $x_{i-1}, \ldots, x_{i-q}$. Being a measure only of the arithmetic operations in the generator rule, the conditional entropy is independent of the lag in the recurrence or the period of the sequence. In that sense it measures a more profound quality of a random number generator than empirical tests with their limited horizon.

Random numbers are the key resource of all Monte Carlo (MC) simulations. They are usually produced by a few lines of code, a subroutine called pseudo random number generator (PRNG). The term *pseudo* refers to the fact these generators implement a *deterministic* recursive formula

$$x_i = f(x_{i-1}, x_{i-2}, \ldots, x_{i-q}), \qquad i > q \qquad (1)$$

to produce a sequence $(x_i)$ of pseudo random numbers (PRNs). The only true randomness in this sequence is concentrated in the choice of the "seed" $(x_1, \ldots, x_q)$, a few hundred bits at most. This small amount of randomness is expanded by (1) to the $10^{10}$ or more random numbers that are consumed by a MC simulation on a present day computer. Most practitioners have no problems founding their scientific reputation on something *pseudo*, they simply trust some well-established PRNG like everybody else trusts the subroutine to calculate $\sin(x)$. Every now and then, however, a popular PRNG is caught producing wrong results, and this does not always entail its removal from the practitioners toolbox. An infamous example is the lagged Fibonacci generator $F(p, q, \circ)$, defined by the recursion

$$x_i = x_{i-p} \circ x_{i-q} \mod m \qquad (2)$$

with $\circ \in \{\oplus, +, -, \times\}$. The bitwise exclusive-or operator $\oplus$ does not require a mod-operation, hence it is very fast. $m$ is usually the word size of the computer or a prime close to it. The choice of the "magic numbers" $p$ and $q$ is based on theoretical considerations, like maximizing the period of the sequence [1]. $F(103, 250, \oplus)$ (also known as R250) has been introduced into the physics community in 1981 [2] as a very fast and reliable PRNG. In 1992, Ferrenberg *et al.* [3] reported serious problems with lagged Fibonacci generators when applied in cluster MC simulations of the 2D-Ising model with the Wolff algorithm. This discovery initiated a series of investigations, in the course of which shortcomings of lagged Fibonacci generators have been found in various other simulations, like in simulations based on the Swendsen-Wang algorithm [4], 3D self-avoiding random walks [5], the Metropolis algorithm on the Blume-Capel model [6], *n*-block tests [7] and 2D-random walks [8]. Despite this bad record, lagged Fibonacci generators kept being recommended as ". . . good enough for many applications" [9] or even for large-scale simulations [10]. The RANLUX generator [11], prevalent in high-energy-physics, is based on a modified $F(p, q, -)$ enhanced with a special measure to improve the quality of the random numbers: it simply throws away up to 80% of the numbers. Other advices to emend lagged Fibonacci generators are to use larger values of the lag like in $F(1393, 4423, \circ)$, or to increase the number of feedback taps to $n > 2$ [8],

$$x_i = x_{i-q_1} \circ x_{i-q_2} \circ \cdots \circ x_{i-q_n} \mod m. \qquad (3)$$

Most of these recommendations are based on empirical evidence only, with the throw-away strategy in RANLUX being a notable exception. But without a theoretical justification these strategies have the smack of sweeping the dirt under the carpet. Some authors blame the 3-point correlations induced by (2) for the problems [6, 12]. In fact, choosing $n > 2$ in (3) the observed deviations are reduced, but strange enough, the simple linear congruence generator

$$x_i = \alpha x_{i-1} \mod m. \qquad (4)$$

with $\alpha \gg 1$ has strong 2-point correlations, yet it performs reasonably in cluster MC simulations. According to Heuer *et al.* [13] ". . . the reason why the Wolff algorithm is so sensitive to triplet correlations remains a mystery." The analysis of a 1D directed random walk simulation by Shchur *et al.* [14, 15] shed light on the mechanism behind this failure: the inability of the PRNG to compensate the persistent bias in the preceding random numbers that is induced by the simulation algorithm. In this contribution we will define a robust, quantitative measure for this inability.

The Wolff algorithm [16, 17] is a very efficient MC method to simulate Ising spin systems in thermal equilibrium. It flips clusters of spins and its central part is the construction of these clusters. The algorithm maintains a list of candidate spins. As long as the candidate list is not empty, one spin is removed from the list and is added to the cluster with probability $P_{\text{add}} = 1 - e^{-2/T}$ where $T$ is the temperature measured in units $k_B/J$. If the spin has been added to the cluster, its equally aligned neighbor spins are added to the candidate list (if they are not themselves part of the cluster). When the candidate list is empty, the growth process stops, all spins of the cluster are flipped and a new growth process is initiated by starting a new cluster with a randomly chosen spin and adding all of its equally aligned neighbors to the candidate list.

---

[*]E-mail:stephan.mertens@physik.uni-magdeburg.de
[†]E-mail:heiko.bauke@physik.uni-magdeburg.de

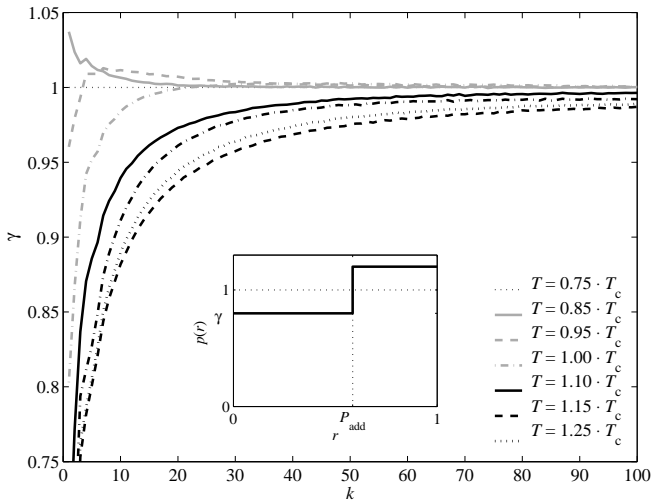

Figure 1: The completion of a cluster in the Wolff algorithm implies a bias $\gamma(k)$ (5) in the preceding random numbers $r_{i-k}$. The bias is towards rejection moves, i.e. the probability of $r_{i-k} < P_{\text{add}}$ (acceptance) is smaller than $P_{\text{add}}$. The inset shows the corresponding probability density of the random numbers $r \in [0,1)$. The data shown is from simulations of a $24 \times 24$ spin square Ising model, but the figures look similar for larger systems and in 3D. $T_c$ is the critical temperature of the infinite system.

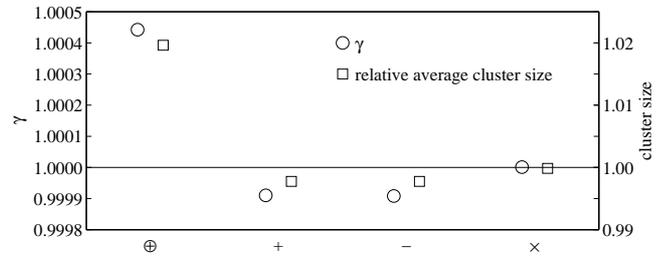

Figure 2: Bias in the output of the lagged Fibonacci generator with biased inputs ($\bigcirc$) and cluster sizes in the Wolff algorithm ($\square$). The bias in the input numbers was $\gamma = 0.975$ and $P_{\text{add}} = 0.586$, corresponding to a MC simulation at the critical temperature of the 2D-Ising model. The simulation was done with $F(13,33,\circ)$ on a $16 \times 16$ spin system. Error bars are smaller than the symbol size. The reference value of the cluster size has been obtained from simulations with a high quality PRNG [18].

In the cluster growth process, the PRNs $r_i = x_i/m \in [0,1)$ are used to decide whether a spin from the candidate list is added to the cluster ($r_i < P_{\text{add}}$) or not ($r_i \geq P_{\text{add}}$). With each spin that is added to the cluster, new spins may enter the candidate list, too. For the growth process to stop, the candidate list needs to be empty, hence we expect the end of each growth process to go hand in hand with a high rejection rate or a super-proportional fraction of random numbers $r_i \geq P_{\text{add}}$. Each time a cluster is completed one can look back at the last random numbers $r_{i-1}, r_{i-2}, \ldots, r_{i-k}, \ldots$ and measure the bias

$$\gamma(k) = \frac{\mathbb{P}(r_{i-k} < P_{\text{add}})}{P_{\text{add}}} \qquad (5)$$

where $\mathbb{P}(r < P)$ denotes the probability of the event $r < P$. Unbiased numbers have $\gamma = 1$, but Fig. 1 shows that the numbers that contribute to the completion of a cluster are indeed strongly biased. For most temperatures the bias is towards rejections moves ($\gamma < 1$). For low temperatures the cluster spans almost the entire system, hence the completion is dominated by a lack of unassigned spins rather than higher rejection rates. In these cases we observe a weak bias towards acceptance moves ($\gamma > 1$).

The bias $\gamma$ is a genuine property of the Wolff algorithm and the PRNG has to cope with this situation: after all we expect it to generate unbiased random numbers $x_i$ even if the numbers $x_{i-1}, \ldots, x_{i-p}$ in (1) are biased. The crucial point is that some PRNGs like the lagged Fibonacci have problems to reconstruct unbiased PRNs from biased input. This is most easily seen for $F(p,q,\oplus)$ and $P_{\text{add}} = 1/2$. Let $P_i$ denote the probability that the most significant bit of $x_i$ equals 1. In the final phase of the cluster growth process we have $P_{i-q} > 1/2$ and $P_{i-p} > 1/2$, and from $x_i = x_{i-q} \oplus x_{i-p}$ we get $P_i < 1/2$, i.e. we expect too many spins to be added to the new cluster.

A simple experiment illustrates the relation between the persistent bias and the cluster size. We draw two real random numbers $r_1$ and $r_2$ from a distribution as shown in the inset of Fig. 1, with bias $\gamma < 1$. These two numbers are used to generate a new random number $r_3$ according to the lagged Fibonacci rule. Fig. 2 shows the bias of $r_3$ for all four binary operators: $\oplus$ leads to a strong bias $\gamma > 1$ (as discussed above), $+$ and $-$ to a weaker bias $\gamma < 1$, and $\times$ shows no bias in the new variable $r_3$. This corresponds nicely with the average cluster size in the Wolff algorithm for simulations of the square Ising model: with $\oplus$ the clusters are too large, with $\pm$ the clusters are too small. Only the $\times$-operator generates clusters of the correct size. These results are consistent with systematic numerical investigations [4].

At this point the virtue of increasing the lag becomes apparent. An increased lag $q$ usually implies an enlarged difference $q - p$, which in turn decreases the probability that *both* numbers $x_{i-p}$ and $x_{i-q}$ are from the completion phase of a cluster. The bias of $x_i$ is much weaker if only one of its predecessors $x_{i-p}$ or $x_{i-q}$ is biased. But he central weakness, the incapacity to restore unbiased PRNs from biased inputs, is independent of the lag. The RANLUX approach to throw away subsets the PRNs helps but requires large fractions of the stream of PRNs to be ignored [19]. To find a better remedy it is instructive to consider a tractable model of a PRNG. Our model PRNG directly generates a stream of real numbers on the interval $[0,1)$ via the recursion

$$r_i = \{\alpha(r_{i-1} + r_{i-2} + \cdots + r_{i-n})\} \qquad (6)$$

where $\{r\}$ denotes the fractional part of $r$. For $\alpha = 1$, Eq. (6)

corresponds to an additive lagged Fibonacci generator with $n$ feedback taps. The case $n = 1$ corresponds to the linear congruence generator (4). Now let $\rho_i$ denote the probability density function of $r_i$. To understand how Eq. (6) transforms the probability density we assume that the input values $r_{i-1}, \ldots, r_{i-n}$ are independent. This holds strictly only for the first iteration on the initial seed, but it allows us to write

$$\rho_i(r) = \frac{1}{\alpha} \sum_{j=0}^{\lfloor n\alpha \rfloor} \rho_{i-1} \star \rho_{i-2} \star \cdots \star \rho_{i-n} \left( \frac{r+j}{\alpha} \right) \qquad (7)$$

for $0 \leq r < 1$. $\lfloor x \rfloor$ denotes the largest integer not larger than $x$ and $\star$ is the convolution operator. The sum results from taking the fractional part in Eq. (6). It can be interpreted as *Riemann sum approximation* to the integral $\int \rho_{i-1} \star \cdots \star \rho_{i-n}(x) \, dx = 1$ with mesh size $1/\alpha$, hence we have the immediate result

$$\lim_{\alpha \to \infty} \rho_i(r) = 1 \qquad 0 \leq r < 1, \qquad (8)$$

i.e. for $\alpha \to \infty$ the new number $r_i$ is uniformly distributed *independently* of the distribution of the preceding numbers. This is what makes the simple linear congruence generator (4) perform better than $F(p,q,\pm)$ and $F(p,q,\oplus)$. One of the numbers $x_{i-q}$ or $x_{i-p}$ that enter the right hand side of the multiplicative generator $F(p,q,\times)$ can be seen as a multiplier for the other. This multiplier varies, but it is $\gg 1$ for almost all iterations. This explains why $F(p,q,\times)$ works fine in cluster MC simulations. For $\alpha = 1$ the support of the $n$-fold convolution is $[0,n)$, and it is sampled with a mesh of size one. In the limit $n \to \infty$ we can again replace the sum by an integral and we get back the uniform distribution for $\rho_i$. For this reason the quality of lagged Fibonacci generators increases with the number of feedback taps. In terms of computational efficiency a generator with a small number of feedback taps but a large factor $\alpha$ is much better than a generator with $\alpha = 1$ but a large number of feedback taps.

We have seen that a multiplier $\alpha > 1$ or a large number $n$ of feedback taps promotes a robust uniformity of the PRNs, even if the input numbers are non uniform. A quantitative measure of this robustness is given by the *entropy* of the output value $r_i$, *conditioned* on the input variables $r_{i-1}, \ldots, r_{i-q}$. Of course $r_i$ is uniquely determined by $r_{i-1}, \ldots, r_{i-q}$ hence the entropy is zero, reflecting the determinism in our PRNs. Fortunately MC simulations use the $r_i$ in a coarse grained manner, to "roll a die" or to "toss a coin". In the Wolff algorithm the (biased) coin shows head with probability $P_{\text{add}}$. In general, a PRNG is used for a random choice out of $M$ of *macrostates* $m_1 \ldots, m_M$ (the faces of the die). Each macrostate $m_j$ is represented by a large number of *microstates*, the actual PRNs. The random choice is done by partitioning the interval $[0,1)$ into disjoint intervals $I_j$ such that $\cup_{j=1}^{M} I_j = [0,1)$, and macrostate $m_j$ is selected if and only if $r \in I_j$. Under the canonical assumption that the PRNs $r$ are uniformly distributed, $m_j$ is selected with probability $P_j = |I_j|$. Obviously a simulation is sensitive only to correlations in the stream $m_j$ of macrostates. On the other hand it can only induce correlations at the macrostate level, and on this level the conditional entropy can be larger than zero: Eq. (1) is deterministic at the microstate level, but it can be chaotic at the macrostate level. The *conditional macrostate entropy H* of a PRNG with $n$ feedback taps reads

$$H = -\sum_{\{m_i\}} \mathbb{P}(m_1, \ldots, m_{n+1}) \log_2 \frac{\mathbb{P}(m_1, \ldots, m_{n+1})}{\mathbb{P}(m_1, \ldots, m_n)}, \qquad (9)$$

where $\mathbb{P}(m_1, \ldots, m_{n+1})$ is the joint probability that the PRNG selects macrostate $m_{n+1}$ and its $n$ input values are *unbiased* and *independent* representatives of the macrostates $m_1, \ldots, m_n$. Note that our definition of $H$ implies maximum entropy of the $n$ microstates that form the input of the generator: $H$ gives the uncertainty that the generator can produce in one iteration, given that we have full knowledge of the preceding macrostates, but no knowledge of the underlying microstates. A good generator should have a value close to the upper bound $-\sum_j P_j \log_2 P_j$. Note also that $H$ is *not* a measure of the stream of PRNs that comes out of a generator. It is a measure of the generator rule itself. Hence it is very different from the entropy used in empirical investigations of pseudo random data [20]. To illustrate this difference consider any empirical test on a a finite stream of pseudo random numbers. The lag $q$ of the underlying generator can easily be tuned to increase the range of the correlations beyond the horizon of the test. The conditional entropy, being independent from the position of the feedback taps, is not deceived by these measures. Increasing the number $M$ of macrostates on the other hand, every PRNG will eventually start to give low entropy values. You better stay away from the microstate level to keep the illusion of randomness.

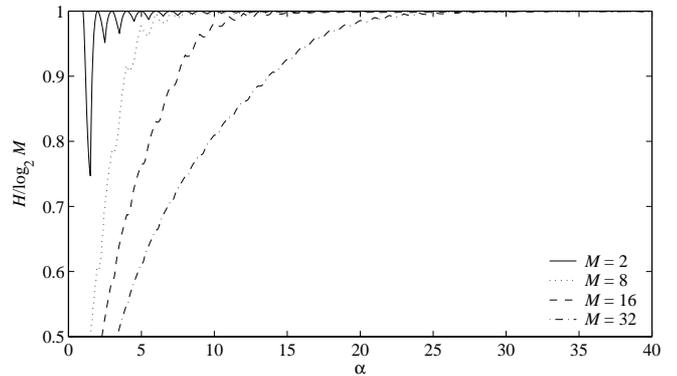

Figure 3: Conditional entropy of the recursion $r_i = \{\alpha(r_{i-p} + r_{i-q})\}$ for different numbers $M$ of equally weighted macrostates. The multiplier $\alpha$ must be larger than $M$ to ensure proper mixing of microstates.

The calculation of the $M^{n+1}$ probabilities that enter the definition of $H$ is straightforward. For simple generators it can be be done analytically. As an example consider $F(p,q,\oplus)$ and $M = 2^m$ equally weighted macrostates: here the macrostate is uniquely determined by the $m$ most significant bits of the $x_i$, and the $\oplus$-operator does not mix these bits with the ambivalent lower order bits. Hence $H = 0$ as opposed to the target value $H = m$. This result is independent of the values of $p$ and $q$ and will be the same for any number of feedback taps. It is an indicator of a fundamental weakness of all $\oplus$-recurrences.

For our model PRNG (6) the probabilities $\mathbb{P}(m_1, \ldots, m_{n+1})$ are simple integrals and computing $H$ is straightforward.

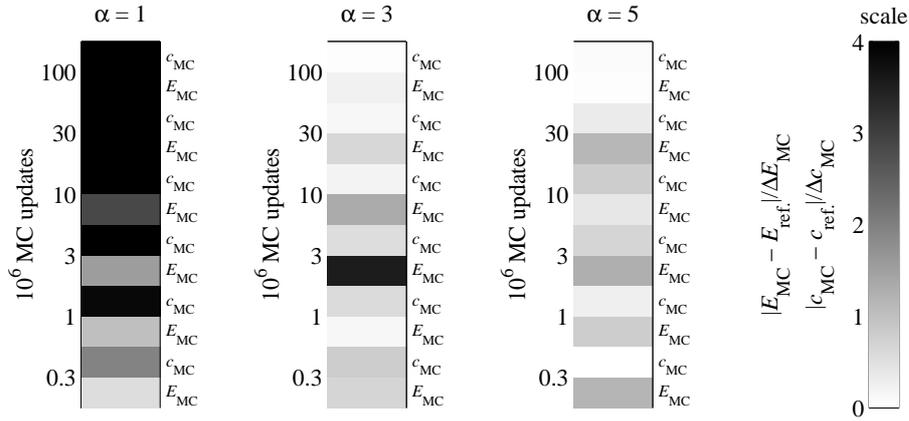

Figure 4: MC simulation of a $6 \times 6 \times 6$ spin Ising model with the Wolff algorithm and $r_i = \alpha(r_{i-13} + r_{i-33}) \mod 2^{31}$ as PRNG. Grayscales indicate the deviation from reference values for the energy $E$ and the specific heat $c$. These reference values have been obtained from simulations with a high quality PRNG [18]. Corresponding simulations with 2D systems yield similar results. For 2D systems reference values of energy and specific heat can be calculated exactly [21].

Fig. 3 shows $H$ for our model PRNG with two feedback taps. The entropy approaches its maximum as the multiplier $\alpha$ gets larger, and for $\alpha > M$ it is very close to the maximum. This is easily understood from Eq. (7): the convolution integral of functions that are constant on intervals of size $1/M$ is well approximated by a Riemann sum of mesh size smaller than $1/M$. Note that $H$ is not a strictly monotonic function of $\alpha$. The resonances in Fig. 3 reflect the fact that even a non uniform distribution at microstate level may yield the correct macrostate statistics for a particular set of weights. According to Fig. 3 even a small factor $\alpha > 1$ yields acceptable statistics for $M = 2$ macrostates. Hence we expect good results in a cluster MC simulation even with a notorious bad generator like $F(p,q,\pm)$ if we enhance the entropy of the latter with a small multiplier $\alpha$. Note that for PRNGs like $F(p,q,\pm)$ the multiplier must be an odd integer. Fig. 4 shows the quality of cluster MC simulations of 2D and 3D Ising models with $F(p,q,+)$ plus multiplier: for $\alpha = 1$ we see strong deviations (which for 2D simulations have long been known), but for $\alpha = 3$, the deviations are much weaker, and for $\alpha = 5$ they have basically disappeared.

### Acknowledgments

All numerical simulations have been done on our Beowulf cluster TINA[22]. This work was supported by Deutsche Forschungsgemeinschaft under grant ME2044/1-1.